%% file: main.tex
\documentclass{article}
\usepackage{spconf}
\ninept
\newcommand{\spaceup}{\vspace{-2.3mm}}
\usepackage{lipsum}

\newcommand\blfootnote[1]{%
  \begingroup
  \renewcommand\thefootnote{}\footnote{#1}%
  \addtocounter{footnote}{-1}%
  \endgroup
}

\input{header.tex}
\DeclareUnicodeCharacter{00A0}{ }
\begin{document}
\topmargin=0mm

 \title{Deep Learning for Joint Source-Channel Coding of Text}

\name{Nariman Farsad$^*$, Milind Rao$^*$\thanks{$^*$The authors contributed equally.}, and Andrea Goldsmith}

\address{Electrical Engineering, Stanford University, Stanford, CA }

\date{}

\maketitle

\begin{abstract}
We consider the problem of joint source and channel coding of structured data such as natural language over a noisy channel. The typical approach to this problem in both theory and practice involves performing source coding to first compress the text and then channel coding to add robustness for the transmission across the channel. This approach is optimal in terms of minimizing end-to-end distortion with arbitrarily large block lengths of both the source and channel codes when transmission is over discrete memoryless channels. However, the optimality of this approach is no longer ensured for documents of finite length and limitations on the length of the encoding. We will show in this scenario that we can achieve lower word error rates  by developing a deep learning based encoder and decoder. While the  approach of separate source and channel coding would minimize bit error rates, our approach preserves semantic information of sentences by first embedding sentences in a semantic space where sentences closer in meaning are located closer together, and then performing joint source and channel coding on these embeddings. 

\blfootnote{This work was funded by the TI Stanford Graduate Fellowship, NSF under
CPS Synergy grant 1330081, and NSF Center for Science of Information
grant NSF-CCF-0939370.} 

\end{abstract}


\spaceup
\section{Introduction}
\spaceup
In digital communications, data transmission typically entails source coding and channel coding. In source coding the data is mapped to a sequence of symbols where the sequence length is optimized. In channel coding redundant symbols are systematically added to this sequence to detect or correct at the receiver the errors that may be introduced during data transfer. One of the consequences of the source-channel coding theorem by Shannon \cite{sha98} is that source and channel codes can be designed separately, with no loss in optimality, for memoryless and ergodic channels when infinite block length codes are used. This is known as the separation theorem, and can be extended to a larger class of channels \cite{vem95sourChan}. 

Optimality of separation in Shannon's theorem assumes no constraint on the complexity of the source and channel code design. However, in practice, having very large block lengths may not be possible due to complexity and delay constraints. Therefore, many communication systems may benefit from designing the source/channel codes jointly. Some examples demonstrating this benefit include: wireless channels \cite{gol95soucChan}, video transmission over noisy channels \cite{zha05jointSCvideo}, and image transmission over noisy channels \cite{dav96jointSCimage,bur13jointSCimage}.

In this work, we consider design of joint source-channel coding for text data with constrained code lengths. Particularly, our ultimate goal is to design a messaging service where sentences are transmitted over an erasure channel. The erasure channel is used here since it can model a broad class of channels where errors are detected but not corrected. One example is timing channels, where information is encoded on the time of release of packets \cite{ana96bitsQueues}. Our proposed coding technique can be used in this channel to create a covert messaging service over packet-switched networks~\cite{dun09secureTiming,kiy13timing,muk16covertTiming,bis2017survey}. In our messaging service, instead of recovering the exact sentence at the receiver, we are interested in recovering the semantic information such as facts or imperatives of the sentence. Therefore, any sentence that conveys the information in the originally transmitted sentence would be considered as an error free output by the decoder even if it differed from the exact sentence. For example, the phrase ``the car stopped'' and ``the automobile stopped'' convey the same information.

One of the first works that considered joint source-channel coding using neural networks is \cite{ron03jointNN}, where simple neural network architectures were used as encoder and decoder for Gauss-Markov sources over additive white Gaussian noise channel. There are also a number of works that use neural networks for compression without a noisy channel (i.e., only source coding). In particular, in \cite{tod16imgComp,tod16fullImgComp} image compression algorithms are developed using RNNs, which outperformed other image compression techniques. Sentence and document encoding is proposed in \cite{li15hierarchical} using neural autoencoders.

{\bf Contributions:} Inspired by the recent success of deep learning in natural language processing for tasks such as machine translation~\cite{wu16GoogleTranslate}, we develop a neural network architecture for joint source-channel coding of text. Our model uses a recurrent neural network (RNN) encoder, a binarization layer, the channel layer, and a decoder based on RNNs. We demonstrate that using this architecture, it is possible to train a joint source-channel encoder and decoder, where the decoder may output a different sentence that preserves its semantic information content. 

We compare the performance of our deep learning encoder and decoder with a separate source and channel coding design\footnote{To the best of our knowledge there are no known joint source-channel coding schemes for text data over erasure channels.}. Since the channel considered here is the erasure channel, we use Reed-Solomon codes for channel coding.  For source coding, we consider three different techniques: a universal source coding scheme, a Huffman coding, and 5-bit character encoding. We demonstrate that the proposed deep learning encoder and decoder outperform the traditional approach in term of word error rate (WER), when the bit budget per sentence encoding is low. Moreover, in many cases, although some words may be replaced, dropped, or added to the sentence by the deep learning decoder, the semantic information in the sentence is preserved in a qualitative sense.




\spaceup
\section{Problem Description} \label{sec:problem_formulation}
\spaceup
In this section, we define our system model associated with  transmitting sentences from a transmitter to a receiver using limited number of bits. 


Let $\mathcal{V}$ be the set of all the words in the vocabulary and let $\vec{s}=[w_1,w_2,\cdots,w_m]$ be the sentence to be transmitted where $w_i\in\mathcal{V}$ is the \ith~word in the sentence. The transmitter converts the sentence into a sequence of bits prior to transmission using source and channel coding. Let $\vec{b} = \varphi_\ell(\vec{s})$ be a binary vector of length-$\ell$, where $\varphi_\ell$ is the function representing the combined effect of the source and channel encoder.  Let $\vec{o}$ be the vector of observations at the receiver corresponding to each of the $\ell$-bits in the transmission. Note that $\vec{o}$ does not necessarily need to be a binary vector, and it could be a vector of real or natural numbers depending on the channel considered.  Let the combined effect of the source and channel decoder function be given by $\nu_\ell(\vec{o})$. Then $\hat{\vec{s}} = [\hat{w}_1,\hat{w}_2, \cdots, \hat{w}_{m^\prime}]=\nu_\ell(\vec{o})$, where 
$\hat{\vec{s}}$ is the recovered sentence.  The traditional approach to designing the source and channel coding schemes is to minimize the word error rate while also minimizing the number of transmission bits. However, jointly optimizing the source coding and the channel coding schemes is a difficult problem  and therefore, in practice, they are treated separately.


The problem considered in this work is designing a joint source-channel coding scheme that preserves the meaning between the transmitted sentence $\vec{s}$ and the recovered sentence $\hat{\vec{s}}$, while the two sentences may have different words and different lengths.

\spaceup
\section{Deep Learning Algorithm}
\begin{figure}
	\centering
	\includegraphics[width=1\columnwidth,keepaspectratio]{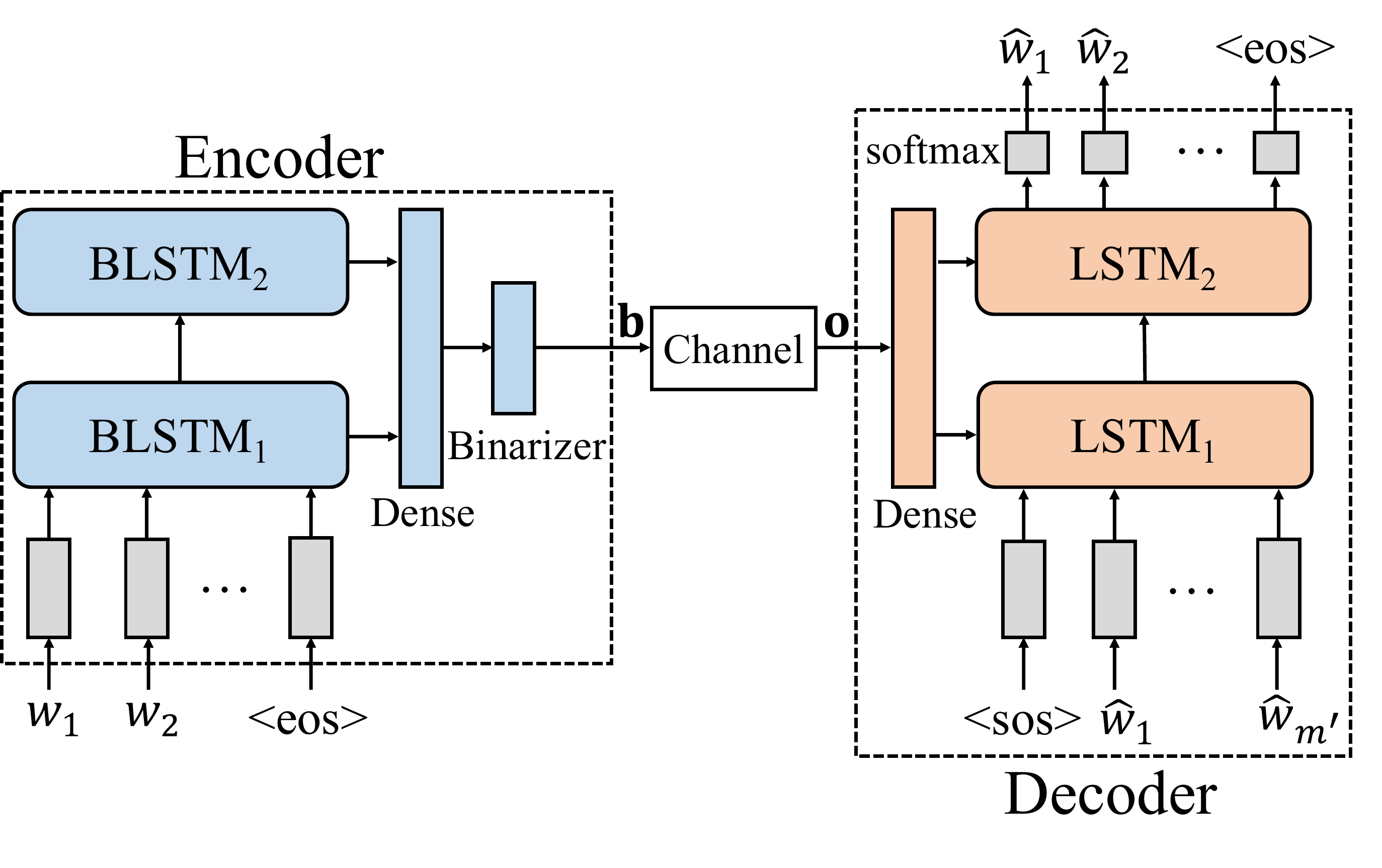}
	\vspace{-0.5cm}
	\caption{\label{fig:encDec} The encoder-decoder architecture.}
	\vspace{-0.5cm}
\end{figure}

\spaceup
\label{sec:algorithms}
Our work is motivated by the recent success of the sequence-to-sequence learning framework \cite{sut14sequence} in different tasks such as machine translation \cite{wu16GoogleTranslate,bah14align}. Our system, which is shown in Fig.~\ref{fig:encDec}, has three components: the encoder, the channel, and the decoder. The encoder takes as an input a sentence $\vec{s}$, concatenated with the special end of sentence word $<$eos$>$, and outputs a bit vector $\vec{b}$ of length $\ell$. The channel takes an input bit vector $\vec{b}$ and produces an output vector $\vec{o}$. The effects of this module is random. The channel output $\vec{o}$ is the input to the decoder, and the output of the decoder is the estimated sentence $\hat{\vec{s}}$. We now describe each of these modules in detail.

\spaceup
\subsection{The Encoder}
The first step in the encoder uses an embedding vector to represent each word in the vocabulary. In this work, we initialize our embedding vectors using Glove \cite{pen14glove}. Let $\vec{E} = [\vec{e}_1, \vec{e}_2, \cdots, \vec{e}_m, \vec{e}_{\text{eos}}]$ be the $m+1$ embeddings of words in the sentence. In the second step, the embedded words are the inputs to a stacked bidirectional long short term memory (BLSTM) network \cite{gra05blstm}. The LSTM cell used in this work is similar to that used in \cite{sak14lstmpeep}. The \jth~BLSTM stack is represented by
\spaceup
\begin{align}
	\vec{C}_j,\vec{H}_j = \blstm_j(\vec{H}_{j-1}),
\end{align} 
where $\vec{C}_j$ is the cell state matrix and $\vec{H}_j$ is output matrix. Each column of $\vec{C}_j$ and $\vec{H}_j$ represents the cell state vector at each time step, and $\vec{H}_0 = \vec{E}$. Fig.~\ref{fig:encDec} shows an encoder with two stacked BLSTM layers.

Let $k$ be the total numbers of BLSTM stacks. We concatenate the outputs at the last step and similarly the cell states at the last step of each layer using \spaceup
\begin{align}
	\vec{h} &= \vec{H}_1[m+1]\oplus\vec{H}_2[m+1]\oplus \cdots \oplus\vec{H}_k[m+1],\\
	\vec{c} &= \vec{C}_1[m+1]\oplus\vec{C}_2[m+1]\oplus \cdots \oplus\vec{C}_k[m+1],
\end{align}
 where $\oplus$ is the concatenation operator, and $\vec{H}_j[m+1]$ and $\vec{C}_j[m+1]$ are the $m+1$ column (i.e., the last step) of, respectively, the outputs and cell states of the \jth~stack.  

To convert $\vec{h}$ and $\vec{c}$ into binary vectors of length $\ell/2$ we use the same technique as in \cite{wil92binarizer,cou15binarizer,tod16imgComp}. The first step in this process uses two fully connected layers \spaceup
\begin{align}
	\vec{h}^* &= \tanh(\vec{W}_h \vec{h}+\vec{a}_h),\\
	\vec{c}^* &= \tanh(\vec{W}_c \vec{c}+\vec{a}_c),
\end{align}
where $\vec{W}_h$ and $\vec{W}_c$ are weight matrices each with $\ell/2$ rows, and $\vec{a}_h$ and $\vec{a}_c$ are the bias vectors. Note that although here we use one fully connected layer, it would be possible to use multiple layers where the size of $\vec{h}$ and $\vec{c}$ is increased or decreased to $\ell/2$ in multiple steps. However, the last layer's activation function must alway be a tanh, to keep the output value in the interval $[-1,1]$. 

The second step maps the the values in $\vec{h}^*$ and $\vec{c}^*$ from the interval $[-1,1]$ to binary values $\{-1, 1\}$. Define a stochastic binarization function as \spaceup
\begin{align}
	\beta(x) = x + Z_x,
\end{align}
where $Z_x \in\{1-x, -x-1\}$ is a random variable distributed according to 	$P(Z_x = 1-x) = \tfrac{1+x}{2}$ and $P(Z_x = -x-1) = \tfrac{1-x}{2}$. Then final binarization step during training is \spaceup
\begin{align}
	\vec{b} = \beta(\vec{h}^*) \oplus \beta(\vec{c}^*)
\end{align}
for the forward pass. During the back-propagation step of the training, the derivative with respect to the expectation $\mathbb{E}[\beta(x)]=x$ is used \cite{rai14binarytech}. Therefore, the gradients pass through the $\beta$ function unchanged.

After training the network using $\beta$, during deployment or testing the stochastic function $\beta(x)$, is replaced with the deterministic function $2u(x)-1$, where $u(x)$ is the unit step function.

\spaceup
\subsection{The Channel}
To allow for end-to-end training of the encoder and the decoder, the channel must allow for back-propagation. Fortunately, some communication channels can be formulated using neural network layers. This includes the additive Gaussian noise channel, multiplicative Gaussian noise channel and the erasure channel. In this work, we consider the erasure channel as it could model packets of data being dropped in a packet switched networks, or wireless channels with deep fades or burst errors. 

The erasure channel can be represented by a dropout layer \cite{sri14dropout}, \spaceup
\begin{align}
	\vec{o} = \text{dropout}(\vec{b},p_d),
\end{align}
where $\vec{o}$ is the vector of observations at the receiver, and $p_d$ is the probability that a bit is dropped. The elements of $\vec{o}$ are in $\{-1,0,1\}$, where 0 indicates erasure (i.e., a dropped bit). Every bit in $\vec{b}$ may be dropped independent of other bits with probability $p_d$.

\spaceup
\subsection{The Decoder}
At the receiver we use a stack of LSTMs for decoding.  The observation vector $\vec{o}$ is input to the decoder.  Let $\ominus(\vec{x},v)$ be the inverse of the concatenation operator, where the vector $\vec{x}$ is broken into $v$ vectors of equal length. Then we have \spaceup
\begin{align}
	\vec{h}^\prime,\vec{c}^\prime = \ominus(\vec{o},2),
\end{align}
which contribute to the initial $\vec{h}^{(j)}_0$ state and $\vec{c}^{(j)}_0$ state of the \jth~ LSTM stack. Particularly, these initial states are given by \spaceup
\begin{align}
	\vec{h}^{(j)}_0 &= \tanh\left(\vec{W}^{(j)}_h\vec{h}^\prime+\vec{a}^{(j)}_h\right),\\
    \vec{c}^{(j)}_0 &= \vec{W}^{(j)}_c\vec{c}^\prime+\vec{a}^{(j)}_c,
\end{align}
where $\vec{W}^{(j)}_h$ and $\vec{W}^{(j)}_c$ are the weight matrix, and $\vec{a}^{(j)}_h$ and $\vec{a}^{(j)}_c$ are the bias vectors.

The first input to the LSTM stack is the embedding vector for a special start of the sentence symbol $<$sos$>$. Note that after the first word $\hat{w}_1$ is estimated, its embedding vector will be used as the input for the next time step. To speed up the training, during the first few epochs, with probability 1 we use the correct word $w_i$ as the input for the $i+1$ time step at the decoder; we gradually anneal the probability with which we replace the correct word $w_i$ with the estimated word $\hat{w}_i$. During deployment and testing we always use the estimated words and the beam search algorithm to find the most likely sequences of words \cite{gra12beamsearch,wu16GoogleTranslate}.

\spaceup
\section{Results}
\label{sec:results}
\spaceup

\begin{figure*}[t!]
	\centering
    \begin{subfigure}[t]{0.5\linewidth}
    \centering
    \includegraphics[width=0.8\linewidth]{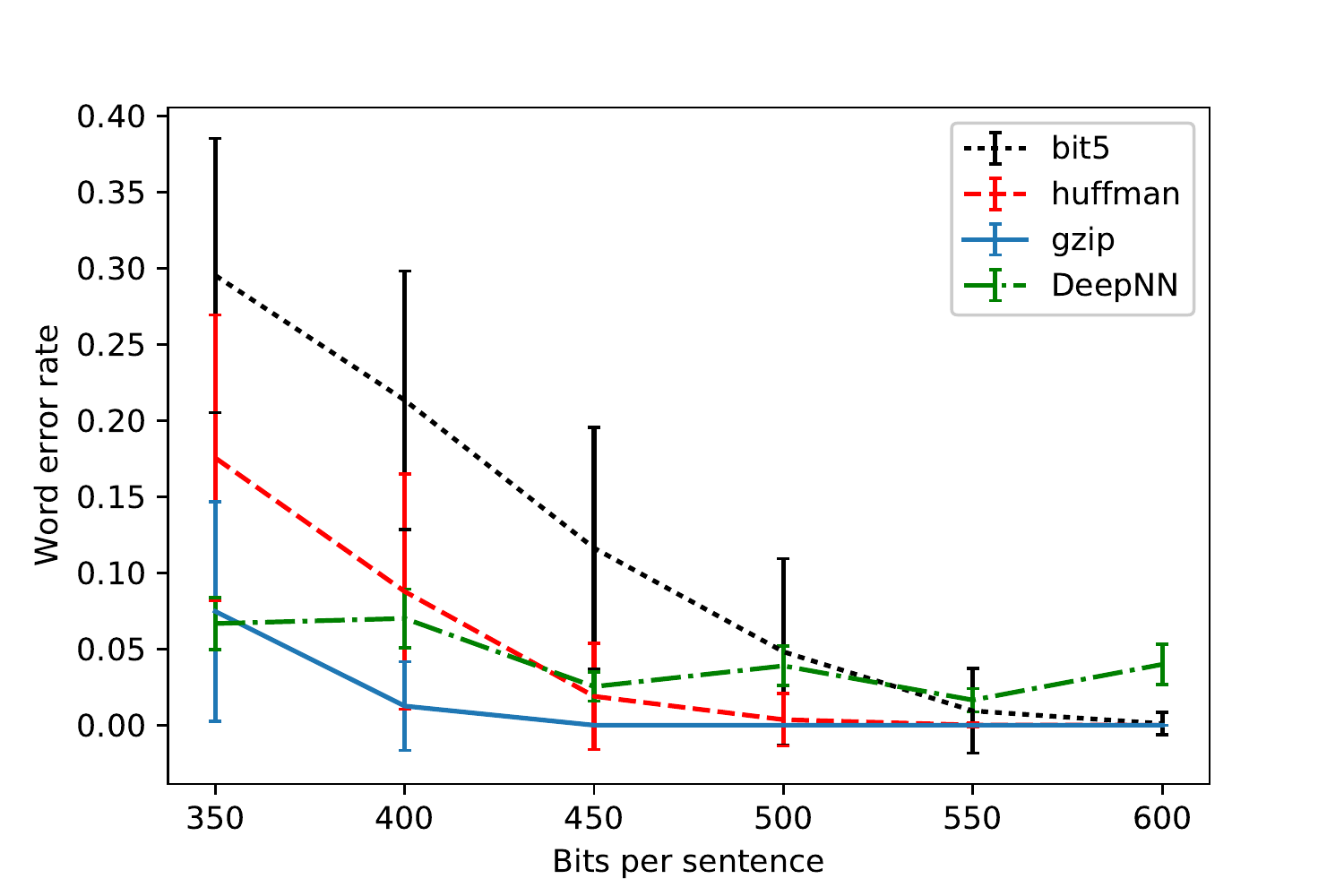}
    \caption{Word error as bits per sentence changes for 0.05 bit erasure probability. \label{fig:bps_error}}
    \end{subfigure}%
    ~
    \begin{subfigure}[t]{0.5\linewidth}
    \centering
    \includegraphics[width=0.8\linewidth]{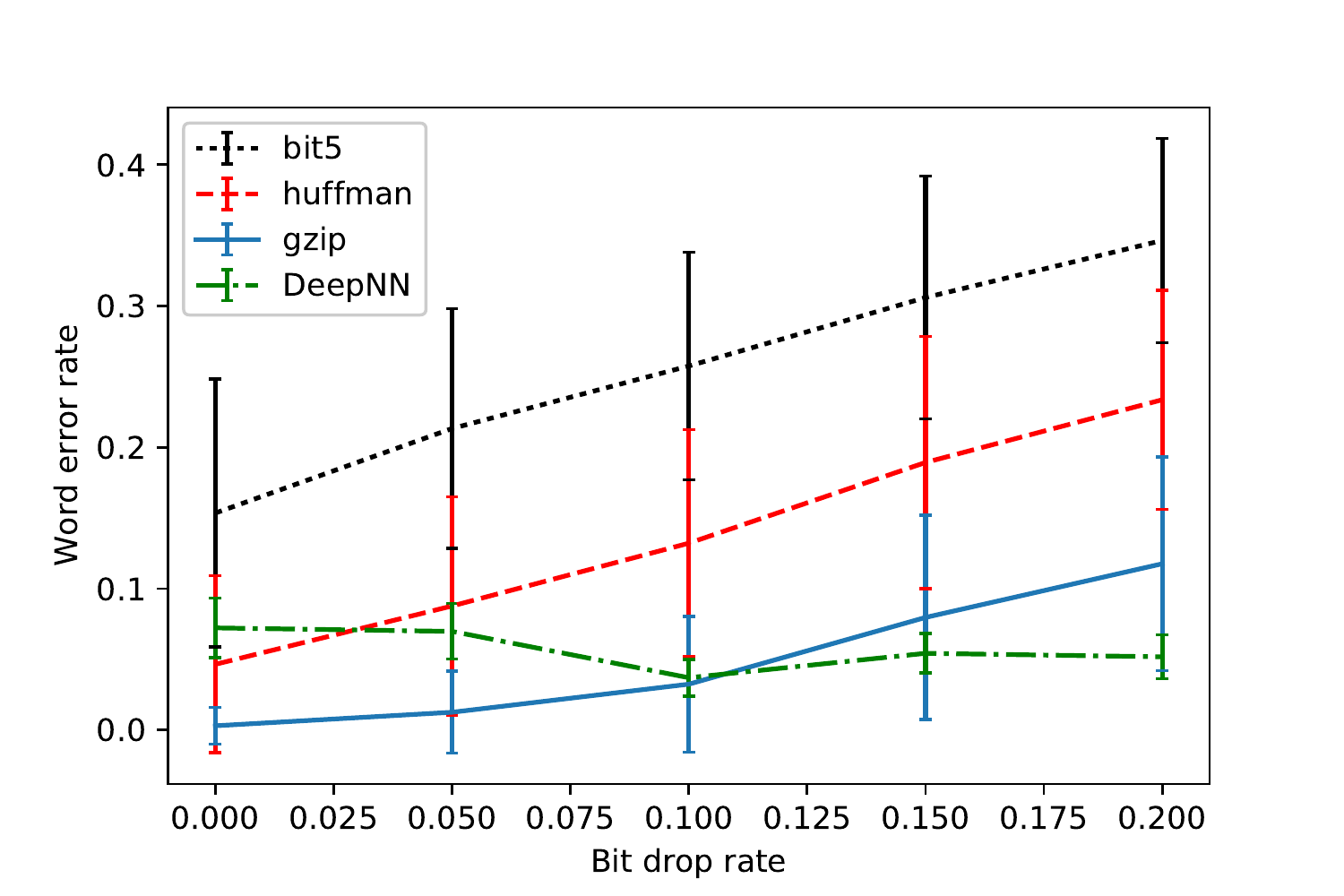}
    \caption{Word error as erasure or bit-drop rate increases for 400 bit encoding. \label{fig:bdr_error}}
    \end{subfigure}%
    
    \begin{subfigure}[t]{0.5\linewidth}
    \centering
    \includegraphics[width=0.8\linewidth]{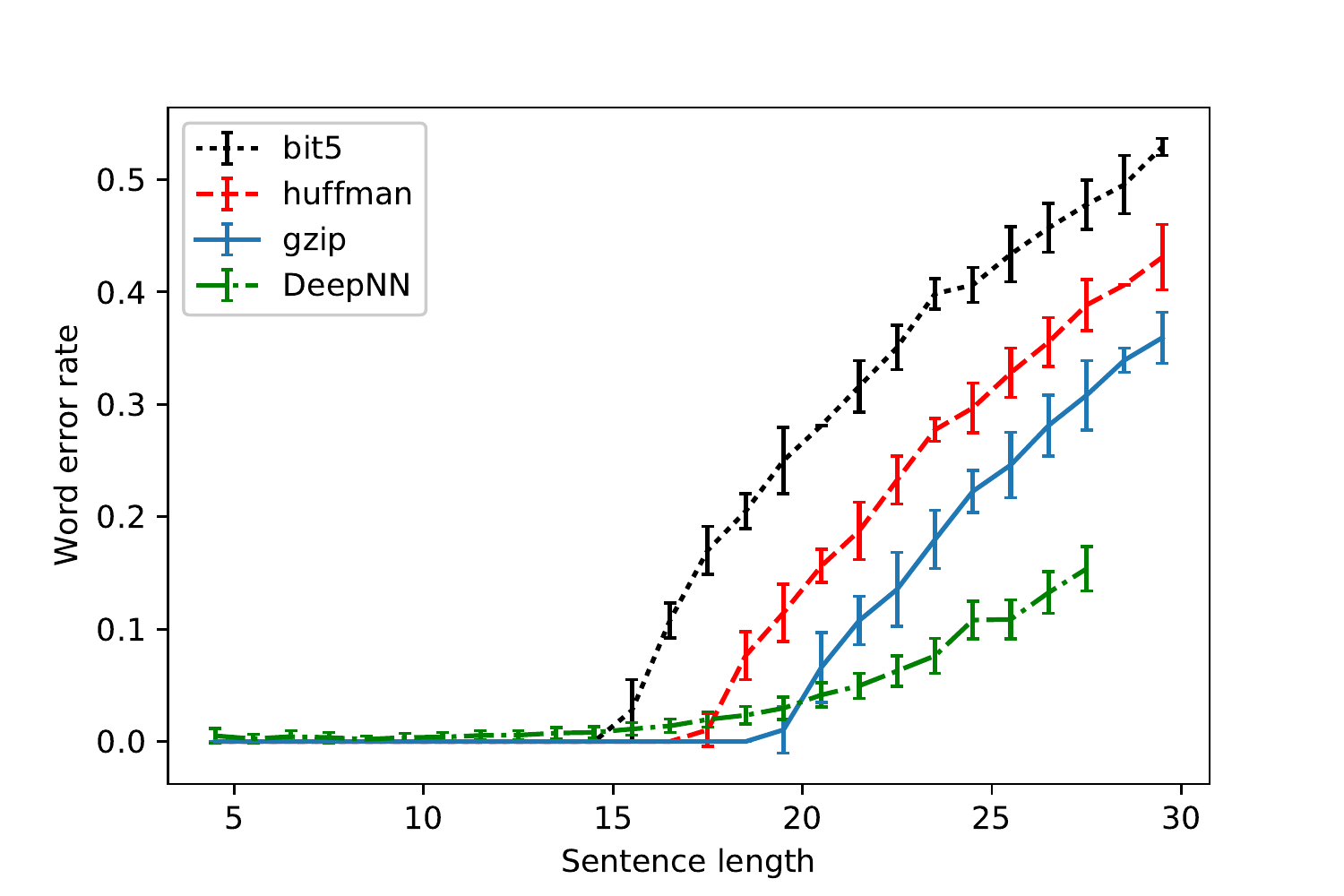}
    \caption{Effect of sentences of different sizes with 400 bit encoding, 0.05 drop rate. \label{fig:sen_len_error}}
    \end{subfigure}%
    ~
    \begin{subfigure}[t]{0.5\linewidth}
    \centering
    \includegraphics[width=0.8\linewidth]{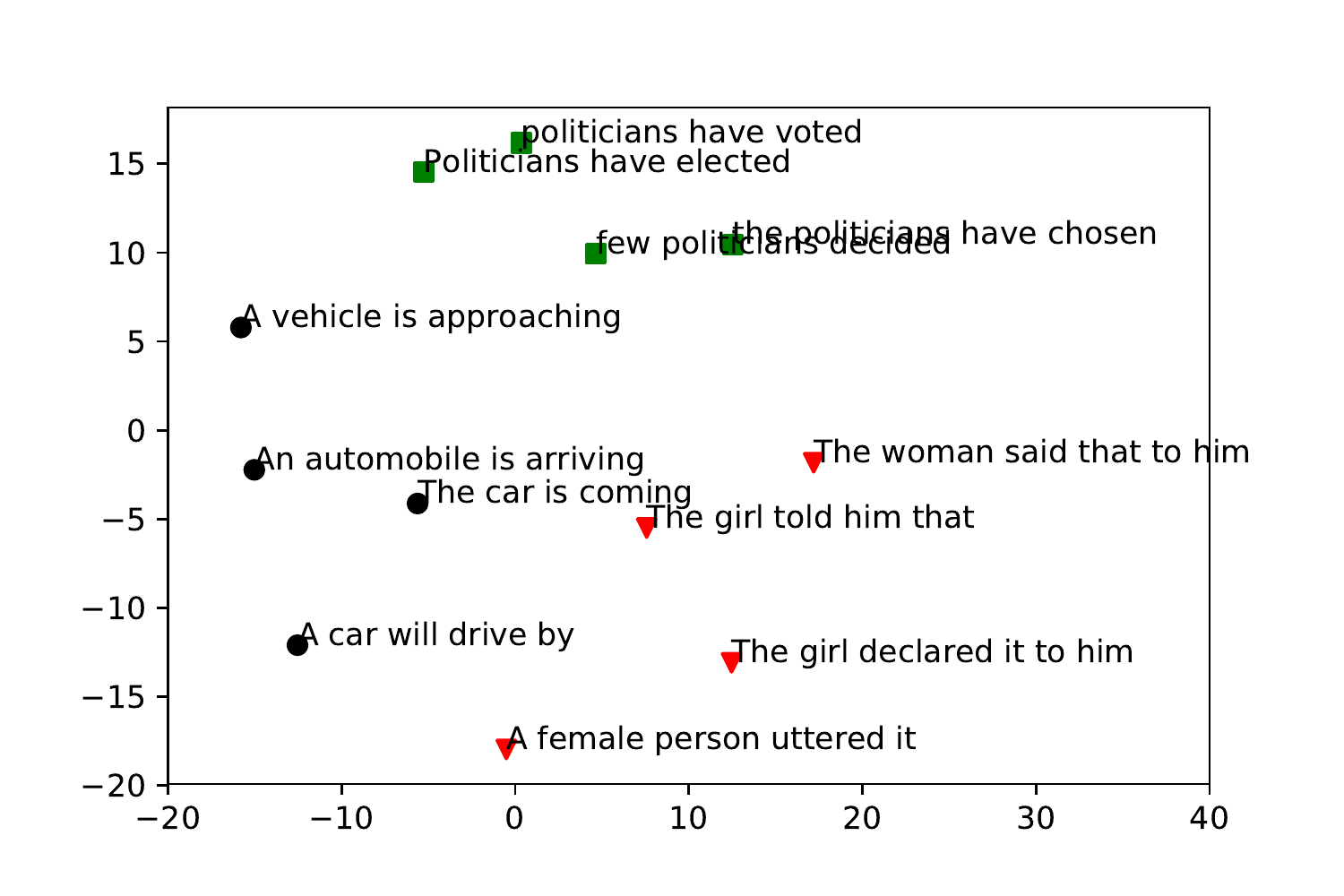}
    \caption{Sample embeddings mapped to two dimensions using manifold dimensions with hamming distances between codes. \label{fig:example_embedding}}
    \end{subfigure}
    \caption{Performance plots.}
    \spaceup
\end{figure*}

\begin{table*}[h!]
  \centering
  \begin{tabular}{|c|p{0.8\linewidth}|}
  \hline
    Punctuation error & TX: efficiency – what efficiency ? \\ 
& RX: efficiency , what efficiency ?   \\ \hline
    Rephrasing & TX: tourism serves as a source of income to totalitarian regimes . \\
& RX: tourism has become a source of income to totalitarian regimes .  \\ \hline
Rephrasing & TX: a few wealthy individuals compared with millions living in hunger . \\
& RX: a few wealthy individuals face with millions living in hunger . \\ \hline
    Tense Error &  TX: a communist country riding roughshod over human rights . \\
& RX: a communist country rides roughshod over human rights .\\ \hline
	An inexplicable error & TX: i listened to colleagues who mentioned bicycles . \\ 
& RX: i listened to colleagues who mentioned goebbels . \\ \hline
Long sentence 1 & TX: there is one salient fact running through these data : the citizens want more information and have chosen television as the best means to receive that information . \\
& RX: there is one glaring weaknesses , by the communication : the citizens want more information and hold ' television as the means to receive this information . \\ \hline
Long sentence 2 & TX: i hope we will be able to provide part - funding for a renovation programme for energy efficiency as a result of this decision of the eu . \\ 
& RX: i hope we will be able to provide for funding for the renovation programme for energy efficiency as a result of decision by the eu . \\ \hline
    \end{tabular}
  \caption{Sample sentences which were transmitted and received using the deep learning approach. \label{tab:samples}}
\end{table*}

In this section, we compare the deep learning approach with traditional information theoretic baselines for bit erasure channels. 

\spaceup
\subsection{The Dataset}
\spaceup 

We work with the proceedings of the European Parliament \cite{koehn2005europarl}. This is a large parallel corpus that is frequently used in statistical machine translation. The English version has around 2.2 million sentences and 53 million words. 

We crawl through the corpus to extract the most common words which we call our vocabulary. We pre-process the dataset by selecting sentences of lengths 4-30 where less then 20\% of the words in the sentences are unknown words (i.e., they are outside of the selected vocabulary). The corpus is split into a training and test data set, where the training set has more that 1.2 million sentences and the test data set has more than 200 thousand sentences. 

\subsection{Deep Learning Approach}
\spaceup
We initialize 200-dimensional word embeddings using the Glove pre-trained embeddings \cite{pen14glove} for words in our vocabulary as well as a few special words (unknowns, padding, start and end symbols). We batch the sentences from the corpus based on their sentence lengths to increase efficiency of computation - i.e.\ sentences of similar length are fed in batches of size 128.

Two layered BLSTM of dimension 256 with peepholes are used for the encoder followed by a dense layer thats brings the dimension of the resultant state to the required bit budget. The decoder has two layers of LSTM cells each with the dimensions 512 with peephole connections. Note that one disadvantage of the deep learning approach is the use of a fixed number of bits for encoding all sentences of different lengths.

\spaceup
\subsection{Separate Source-Channel Coding Baselines}
\spaceup
We implement separate source and channel coding which we know is optimal in the asymptote of arbitrarily large block lengths and delays. The source coding is done using three approaches:
\begin{enumerate}
\item Universal compressors: We use gzip which combines a Lempel-Ziv universal compression \cite{ziv1977universal} scheme with Huffman coding. This method works universally with all kinds of data and theoretically reaches the entropy limit of compression in the asymptote. However, since this technique does not work well for single sentences, we improve its performance by jointly compressing sentences in batches of size 32 or more. Note that this will give this technique an unfair advantage since it will no longer perform source coding on single sentences.

\item Huffman coding: To allow for single sentence source coding, we use Huffman coding on characters in the sentence. Using the training corpus, we compute character frequencies, which are then used to generated the Huffman codebook. 

\item Fixed length character encoding: In this approach, we use a fixed 5-bit encoding for characters (the corpus is converted to lower case) and some special symbols. Decoding gzip and Huffman codes when there are errors or corruptions in the output of the channel decoder is not trivial. However, this baseline with 5-bit encoding can be decoded.
\end{enumerate}

After source encoding using the above approaches, we use a Reed-Solomon code \cite{reed1960polynomial} that can correct up to the expected number of erasures. In the comparison, we assume the channel code can exactly compensate for erasures that occur. This assumption favors separate source-channel coding baselines as we can expect the number of bit erasures to be larger than the expected number with high probability. If this occurs, the channel decoding process will have errors and this may result in irredeemable corruption for decoding the source codes (gzip or huffman). 

Finally, we compare performance by using a fixed bit budget per sentence. However, these schemes inherently produce embeddings of different lengths. If the encoding of a sentence exceeds the bit budget, we re-encode the sentence without its last word (resulting in a word error). We repeat the procedure until the encoding is within the bit limit. 

\spaceup
\subsection{Performance}
\spaceup 
There is no better metric than a human judge to establish the similarity between sentences. As a proxy, we measure performance of the deep learning approach as well as the baselines using the edit distance or the Levenshtein distance. This metric is commonly used to measure the dissimilarity of two strings. It is computed using a recursive procedure that establishes the minimum number of word insertion, deletion, or substitution operations that would transform one sentence to another. The edit distance normalized by the length of the sentence is what we refer to as the word error rate. Word error rate is commonly used to evaluate performance in speech recognition and machine translation \cite{quirk2004monolingual,wubben2012sentence}. A downside of the metric is that it cannot capture the effect of synonyms or other aspects of semantic similarity. 

In Fig.\ \ref{fig:bps_error}, we study the impact of the bit budget or the number of bits per sentence on the word error rate when we have a bit erasure probability of 0.05. Among the traditional baselines, gzip outperforms Huffman codes, and Huffman codes outperform the fixed length encoding. All three approaches result in no error if the bit allocation exceeds the number of bits required. This is because we assume the Reed-Solomon code compensates for all channel erasures. We observe that the deep learning approach is most competitive with limited bit allocations. As we enter the regime of excessive redundancy, the word error rate continually falls. 

In Fig.\ \ref{fig:bdr_error}, we look at the impact of the channel on word error rates when we have a bit allocation of 400 bits per sentence. Between the traditional baselines, we observe again that gzip is optimal as it operates on large batches followed by Huffman codes. 400 bits is not enough to completely encode sentences even when the channel is lossless. We make the observation again that in stressed environments (low bit allocations for large bit erasure rates), the deep learning approach outperforms the baselines. 

What Fig.\ \ref{fig:bps_error} and Fig.\ \ref{fig:bdr_error} hide is the impact of varying sentence lengths. If we consider a batch of sentences in random order from the corpus, we will have both large and short sentences. Traditional baselines can allot large encodings to long sentences and short encodings to others leading to an averaged bit allocation that may be short with few errors. However, the deep learning approach has the same bit allocation for sentences regardless of their length. We can improve the performance of the deep learning approach here by varying the length of the embedding based on the sentence length. 

Fig.\ \ref{fig:sen_len_error} illustrates this very clearly. In this case, instead of having batches with sentences of different lengths, we use homogeneous batches to show the impact of the sentence lengths on word error rates (bit allocation 400, bit erasure rate 0.05). For short sentences, we are in the excess bit allocation regime. As the sentence length increases beyond 20, the deep learning approach significantly outperforms baselines. Another aspect to consider is that word errors of the deep learning approach may not be word errors - that may include substitutions of words using synonyms or rephrasing which does not change the meaning of the word. 

\spaceup
\subsection{Properties of the encoding}
\spaceup 

The deep learning approach results in a lossy compression of text. It is able to do this by encoding a semantic embedding of the sentence. We can watch this in action in Fig.\ \ref{fig:example_embedding}. Here, we compute the embeddings of a few sentences, groups of which are thematically linked. One group of sentences is about a girl saying something to a man, another is about a car driving and the last is about politicians voting. We then find the Hamming distance between the embeddings and use this dissimilarity matrix and multidimensional scaling approaches \cite{borg2005modern} to view it in two dimensions. Sentences that express the same idea have embeddings that are close together in Hamming distance. We do not see such behavior in information theoretic baselines which do not consider the fact that it is text with semantic information that they are encoding.

A few representative errors are shown in Table \ref{tab:samples}.

\spaceup
\section{Conclusion}
\label{sec:conclusion}
\spaceup

We considered the problem of joint source-channel coding of text data using deep learning techniques from natural language processing. For example, in many applications, recovery of the exact transmitted sentence may not be important as long as the main information within the sentence is conveyed. We demonstrated that our proposed joint source-channel coding scheme outperforms separate source and channel coding, especially in scenarios with a small number of bits to describe each sentence. 

One drawback of the current algorithm is that it uses a fixed bit length to encode sentences of different length. As part of future work, we investigate how to resolve this issue. With severe bit restrictions per sentence, we will also look at deep learning based summarization to represent information. Joint source-channel coding of other forms of structured data such as images, audio, and video would also be a relevant future direction.

\bibliographystyle{IEEEbib}
\bibliography{ICASSP18}

\end{document}

%% file: header.tex
\usepackage[utf8]{inputenc}
\usepackage[cmex10]{amsmath}
\usepackage{amssymb}
\usepackage{amsthm}
\usepackage{wasysym}
\usepackage{balance}
\usepackage{url}

\usepackage{graphicx}
\usepackage{epstopdf}
\usepackage{tikz}
\usepackage{tikz-3dplot}
\usetikzlibrary{fit}
\usepackage{pgfplots}
 \usepgfplotslibrary{groupplots}
 \usepgfplotslibrary{external}
\usepackage{textcomp}

\usepackage{color}
\usepackage{subcaption}
\usepackage{bm}

\def\ith{$i^{\text{th}}$}	
\def\jth{$j^{\text{th}}$}
\renewcommand{\vec}[1]{\mathbf{#1}}

\newcommand{\blstm}{\mathrm{BLSTM}}

\newcommand{\be}{\begin{equation}}

\newcommand{\bc}{\begin{center}}
\newcommand{\bfl}{\begin{flushleft}}

\newcommand{\beqa}{\begin{eqnarray}}
\newcommand{\beqan}{\begin{eqnarray*}}
\newcommand{\beq}{\begin{equation}}

\renewcommand*\env@matrix[1][*\c@MaxMatrixCols c]{%
  \hskip -\arraycolsep
  \let\@ifnextchar\new@ifnextchar
  \array{#1}}
  
\usepackage{etoolbox}
\preto{\section}{\setcounter{subsubsection}{0}}

%% file: main.bbl
\begin{thebibliography}{10}

\bibitem{sha98}
Claude~E Shannon and Warren Weaver,
\newblock {\em The mathematical theory of communication},
\newblock University of Illinois press, 1998.

\bibitem{vem95sourChan}
Sridhar Vembu, Sergio Verdu, and Yossef Steinberg,
\newblock ``The source-channel separation theorem revisited,''
\newblock {\em IEEE Transactions on Information Theory}, vol. 41, no. 1, pp.
  44--54, 1995.

\bibitem{gol95soucChan}
A.~Goldsmith,
\newblock ``Joint source/channel coding for wireless channels,''
\newblock in {\em IEEE Vehicular Technology Conference. Countdown to the
  Wireless Twenty-First Century}, Jul 1995, vol.~2, pp. 614--618 vol.2.

\bibitem{zha05jointSCvideo}
Fan Zhai, Yiftach Eisenberg, and Aggelos~K Katsaggelos,
\newblock ``Joint source-channel coding for video communications,''
\newblock {\em Handbook of Image and Video Processing}, 2005.

\bibitem{dav96jointSCimage}
Geoffrey Davis and John Danskin,
\newblock ``Joint source and channel coding for image transmission over lossy
  packet networks,''
\newblock in {\em Conf. Wavelet Applications to Digital Image Processing},
  1996, pp. 376--387.

\bibitem{bur13jointSCimage}
Ozgun~Y Bursalioglu, Giuseppe Caire, and Dariush Divsalar,
\newblock ``Joint source-channel coding for deep-space image transmission using
  rateless codes,''
\newblock {\em IEEE Transactions on Communications}, vol. 61, no. 8, pp.
  3448--3461, 2013.

\bibitem{ana96bitsQueues}
Venkat Anantharam and Sergio Verdu,
\newblock ``Bits through queues,''
\newblock {\em IEEE Transactions on Information Theory}, vol. 42, no. 1, pp.
  4--18, 1996.

\bibitem{dun09secureTiming}
Brian~P Dunn, Matthieu Bloch, and J~Nicholas Laneman,
\newblock ``Secure bits through queues,''
\newblock in {\em IEEE Information Theory Workshop}. IEEE, 2009, pp. 37--41.

\bibitem{kiy13timing}
Negar Kiyavash, Farinaz Koushanfar, Todd~P Coleman, and Mavis Rodrigues,
\newblock ``A timing channel spyware for the csma/ca protocol,''
\newblock {\em IEEE Transactions on Information Forensics and Security}, vol.
  8, no. 3, pp. 477--487, 2013.

\bibitem{muk16covertTiming}
Pritam Mukherjee and Sennur Ulukus,
\newblock ``Covert bits through queues,''
\newblock in {\em IEEE Conference on Communications and Network Security
  (CNS)}. IEEE, 2016, pp. 626--630.

\bibitem{bis2017survey}
Arnab~Kumar Biswas, Dipak Ghosal, and Shishir Nagaraja,
\newblock ``A survey of timing channels and countermeasures,''
\newblock {\em ACM Computing Surveys (CSUR)}, vol. 50, no. 1, pp. 6, 2017.

\bibitem{ron03jointNN}
Li~Rongwei, Wu~Lenan, and Guo Dongliang,
\newblock ``Joint source/channel coding modulation based on bp neural
  networks,''
\newblock in {\em Proceedings of the International Conference on Neural
  Networks and Signal Processing}. IEEE, 2003, vol.~1, pp. 156--159.

\bibitem{tod16imgComp}
George~Toderici et~al.,
\newblock ``Variable rate image compression with recurrent neural networks,''
\newblock in {\em International Conference on Learning Representations}, 2016.

\bibitem{tod16fullImgComp}
George Toderici, Damien Vincent, Nick Johnston, Sung~Jin Hwang, David Minnen,
  Joel Shor, and Michele Covell,
\newblock ``Full resolution image compression with recurrent neural networks,''
\newblock {\em arXiv preprint arXiv:1608.05148}, 2016.

\bibitem{li15hierarchical}
Jiwei Li, Minh-Thang Luong, and Dan Jurafsky,
\newblock ``A hierarchical neural autoencoder for paragraphs and documents,''
\newblock {\em arXiv preprint arXiv:1506.01057}, 2015.

\bibitem{wu16GoogleTranslate}
Yonghui~Wu et~al.,
\newblock ``Google's neural machine translation system: Bridging the gap
  between human and machine translation,''
\newblock {\em CoRR}, vol. abs/1609.08144, 2016.

\bibitem{sut14sequence}
Ilya Sutskever, Oriol Vinyals, and Quoc~V Le,
\newblock ``Sequence to sequence learning with neural networks,''
\newblock in {\em Advances in neural information processing systems}, 2014, pp.
  3104--3112.

\bibitem{bah14align}
Dzmitry Bahdanau, Kyunghyun Cho, and Yoshua Bengio,
\newblock ``Neural machine translation by jointly learning to align and
  translate,''
\newblock {\em arXiv preprint arXiv:1409.0473}, 2014.

\bibitem{pen14glove}
Jeffrey Pennington, Richard Socher, and Christopher Manning,
\newblock ``Glove: Global vectors for word representation,''
\newblock in {\em Proceedings of the 2014 conference on empirical methods in
  natural language processing (EMNLP)}, 2014, pp. 1532--1543.

\bibitem{gra05blstm}
Alex Graves and J{\"u}rgen Schmidhuber,
\newblock ``Framewise phoneme classification with bidirectional lstm and other
  neural network architectures,''
\newblock {\em Neural Networks}, vol. 18, no. 5, pp. 602--610, 2005.

\bibitem{sak14lstmpeep}
Ha{\c{s}}im Sak, Andrew Senior, and Fran{\c{c}}oise Beaufays,
\newblock ``Long short-term memory recurrent neural network architectures for
  large scale acoustic modeling,''
\newblock in {\em Fifteenth Annual Conference of the International Speech
  Communication Association}, 2014.

\bibitem{wil92binarizer}
Ronald~J Williams,
\newblock ``Simple statistical gradient-following algorithms for connectionist
  reinforcement learning,''
\newblock {\em Machine learning}, vol. 8, no. 3-4, pp. 229--256, 1992.

\bibitem{cou15binarizer}
Matthieu Courbariaux, Yoshua Bengio, and Jean-Pierre David,
\newblock ``Binaryconnect: Training deep neural networks with binary weights
  during propagations,''
\newblock in {\em Advances in Neural Information Processing Systems}, 2015, pp.
  3123--3131.

\bibitem{rai14binarytech}
Tapani Raiko, Mathias Berglund, Guillaume Alain, and Laurent Dinh,
\newblock ``Techniques for learning binary stochastic feedforward neural
  networks,''
\newblock {\em stat}, vol. 1050, pp. 11, 2014.

\bibitem{sri14dropout}
Nitish Srivastava, Geoffrey~E Hinton, Alex Krizhevsky, Ilya Sutskever, and
  Ruslan Salakhutdinov,
\newblock ``Dropout: a simple way to prevent neural networks from
  overfitting.,''
\newblock {\em Journal of machine learning research}, vol. 15, no. 1, pp.
  1929--1958, 2014.

\bibitem{gra12beamsearch}
Alex Graves,
\newblock ``Sequence transduction with recurrent neural networks,''
\newblock {\em arXiv preprint arXiv:1211.3711}, 2012.

\bibitem{koehn2005europarl}
Philipp Koehn,
\newblock ``Europarl: A parallel corpus for statistical machine translation,''
\newblock in {\em MT summit}, 2005, vol.~5, pp. 79--86.

\bibitem{ziv1977universal}
Jacob Ziv and Abraham Lempel,
\newblock ``A universal algorithm for sequential data compression,''
\newblock {\em IEEE Transactions on information theory}, vol. 23, no. 3, pp.
  337--343, 1977.

\bibitem{reed1960polynomial}
Irving~S Reed and Gustave Solomon,
\newblock ``Polynomial codes over certain finite fields,''
\newblock {\em Journal of the society for industrial and applied mathematics},
  vol. 8, no. 2, pp. 300--304, 1960.

\bibitem{quirk2004monolingual}
Chris Quirk, Chris Brockett, and William Dolan,
\newblock ``Monolingual machine translation for paraphrase generation,''
\newblock in {\em Proceedings of the 2004 conference on empirical methods in
  natural language processing}, 2004.

\bibitem{wubben2012sentence}
Sander Wubben, Antal Van Den~Bosch, and Emiel Krahmer,
\newblock ``Sentence simplification by monolingual machine translation,''
\newblock in {\em Proceedings of the 50th Annual Meeting of the Association for
  Computational Linguistics: Long Papers-Volume 1}. Association for
  Computational Linguistics, 2012, pp. 1015--1024.

\bibitem{borg2005modern}
Ingwer Borg and Patrick~JF Groenen,
\newblock {\em Modern multidimensional scaling: Theory and applications},
\newblock Springer Science \& Business Media, 2005.

\end{thebibliography}
